# Biomolecular NMR at 1.2 GHz

Lucia Banci[1,2]*, Letizia Barbieri[1], Vito Calderone[1,2], Francesca Cantini[1,2], Linda Cerofolini[1], Simone Ciofi-Baffoni[1,2], Isabella C. Felli[1,2], Marco Fragai[1,2], Moreno Lelli[1,2], Claudio Luchinat[1,2]*, Enrico Luchinat[1,3], Giacomo Parigi[1,2], Mario Piccioli[1,2], Roberta Pierattelli[1,2], Enrico Ravera[1,2], Antonio Rosato[1,2], Leonardo Tenori[1], Paola Turano[1,2]

[1] Magnetic Resonance Center (CERM), University of Florence, and Consorzio Interuniversitario Risonanze Magnetiche di Metalloproteine (CIRMMP), via L. Sacconi 6, 50019 Sesto Fiorentino, Italy;

[2] Department of Chemistry "Ugo Schiff", University of Florence, via della lastruccia 3, 50019 Sesto Fiorentino, Italy

[3] Department of Experimental and Clinical Biomedical Sciences "Mario Serio", University of Florence, viale G.B. Morgagni 50, 50134 Firenze, Italy

*Correspondance to: Lucia Banci (banci@cerm.unifi.it) and Claudio Luchinat (luchinat@cerm.unifi.it)

Abstract: The development of new superconducting ceramic materials, which maintain the superconductivity at very intense magnetic fields, has prompted the development of a new generation of highly homogeneous high field magnets that has trespassed the magnetic field attainable with the previous generation of instruments. But how can biomolecular NMR benefit from this? In this work, we review a few of the notable applications that, we expect, will be blooming thanks to this newly available technology.

Introduction

Nuclear magnetic resonance spectroscopy is a key enabling methodology in several areas of modern chemistry, because of its privileged look at the atomic level of molecules. Biomolecular applications are one of the most common flavors of NMR, where our laboratory has specialized over more than four decades. In spite of its broad applicability and power, NMR is strongly limited by sensitivity, because of the limited energy difference between the energy states of the nuclear spins. The energy difference can be increased by increasing the static magnetic field. Over the years, the development of superconducting alloys that maintain a high critical current even at high magnetic fields had led to produce commercially available magnets with fields up to 23.5 T, which is the limit of the metallic alloy technology. The discovery of superconductive ceramic oxides, which maintain a very high critical current in fields up to 40 T, has triggered the development of a new generation of commercial instruments.

The increase in attainable fields brings with it an obvious increase in price tag, and since the early development of NMR the question has always been raised whether purchasing top of the range instruments would still provide good value for money. The question became dramatic when top of the range instruments moved (approximately) from 800 to 900 MHz (Figure 1), but the answer from the scientific community was undoubtedly positive. Figure 1 is even more encouraging with respect to the future, because it shows that the price per MHz is rather levelling off.

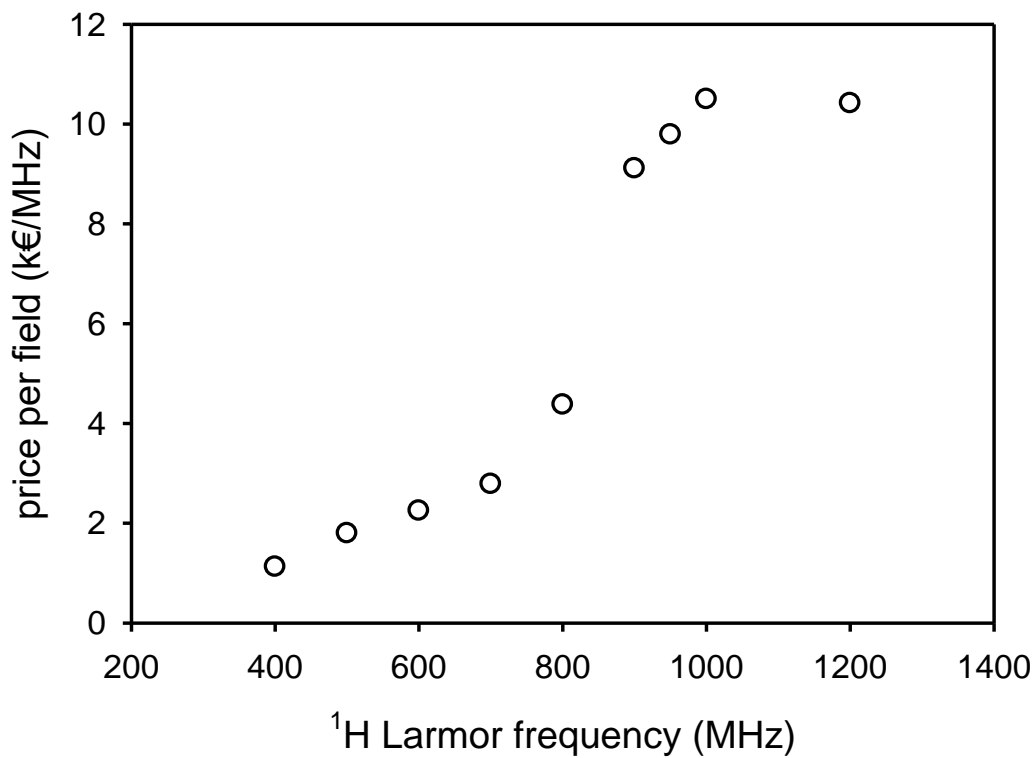

Figure 1. Trend of the price of commercial NMR spectrometers using superconducting magnet systems normalized by the field, at the time of market introduction of the 1.2 GHz instrument (source: Bruker Biospin).

Our laboratory will soon be equipped with the first 28.2 T (1.2 GHz) magnet system. In this review, we show some of the first spectra obtained on our instrument while still under factory tests, demonstrating that this new generation of magnets will have a significant impact on the development of biomolecular NMR.

Materials and methods

Samples

$^{13}$C,$^{15}$N α-synuclein was prepared as previously reported.(1) The NMR sample was 0.9 mM in 20 mM phosphate buffer at pH 6.5, 100 mM NaCl, 50 μM EDTA, in water with 5% D$_2$O for the lock, in a 3 mm NMR tube.

The double mutant (H3N, H4N) of human carbonic anhydrase II (HCAII) was used for the acquisition of 2D $^1$H-$^{15}$N $^1$H-detected and $^{15}$N-detected TROSY-HSQC experiments and as a model system to investigate the effect of parmagnetism. $^{15}$N carbonic anhydrase II was prepared as previously reported.(2) The cobalt(II) and zinc(II) derivatives (ZnHCAII and CoHCAII) were obtained by a demetalation/metallation approach.(3) The NMR samples were 0.6 mM in 10 mM HEPES buffer at pH 5.8 with 0.625 mM oxalate, in water with 10% D$_2$O for the lock, in a 3 mm NMR tube for the experiments at 1.2 GHz and in a 5 mm tube for the experiments at 950 MHz.

Wild-type HCAII was instead used in the investigation of sensitivity and resolution in diluted systems. The NMR sample was 100 μM in 4 mM sodium/potassium phosphate buffer at pH 7.4 with 155 mM NaCl in water with 10% D$_2$O for the lock, in a 3 mm NMR tube.

The non-glycosylated antibody is a 145 kDa protein in isotopic natural abundance. The NMR sample was 10 mg/mL in 10 mM citric acid monohydrate buffer at pH 6.0, 50 mM NaCl, 150 mM Sucrose, 1% (w/v) Tween 80 in water with 10% D$_2$O for the lock, in a 3 mm NMR tube.

Spectra at 1.2 GHz were recorded on a Bruker Avance NEO spectrometer operating at 1.2 GHz with a 28.2 T HTS/LTS hybrid magnet. The instrument is equipped with a 3 mm, triple resonance TCI cryo-probehead.

Spectra for comparison were recorded on Bruker Avance NEO or Avance III HD spectrometers operating at various frequencies. Specifications are given in each section.

Results and discussion

The TROSY effect and nitrogen detection

Since the TROSY effect was observed, it has been clear that the optimal field to achieve the maximum of the interference between dipolar and CSA relaxation mechanisms, hence the maximum value of the lifetime of single quantum coherences, is around 900 MHz.(4) Considering the fact that the efficiency of the inductive detection increases with the magnetic field to the power of 3/2,(5) the signal-to-noise ratio per unit of time is expected to increase greatly on moving to higher fields.(6) Comparing the TROSY spectra acquired at 1200 MHz and 950 MHz on a sample of a 30 KDa zinc-enzyme, human carbonic anhydrase II (ZnHCAII), it can be observed that the linewidths of the proton resonances decrease when expressed in ppm (Figure 2).

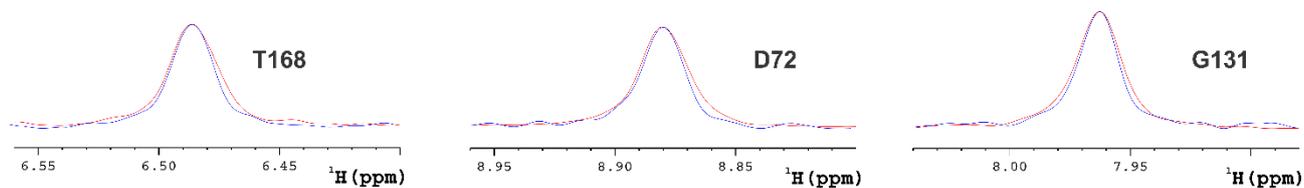

Figure 2. Slices of the $^1$H-detected 2D $^1$H-$^{15}$N TROSY spectra recorded on 950 Bruker AVANCEIIIHD (red) and 1200 Bruker AVANCE NEO (blue) spectrometers on a sample of ZnHCAII. The 950 MHz spectrometer is equipped with a 5 mm TCI 3 channels HCN cryo-probehead, while the 1200 MHz spectrometer is equipped with a 3 mm TCI 3 channels HCN cryo-probehead. The spectra have been processed using the same number of points per ppm. The signals appear sharper in the spectrum acquired at 1200 MHz.

Even more so, as already proposed,(6) a further advantage of the higher fields is manifested in the detection of heteronuclei with lower gyromagnetic ratios, which becomes a captivating option at the highest fields (Figure 3).

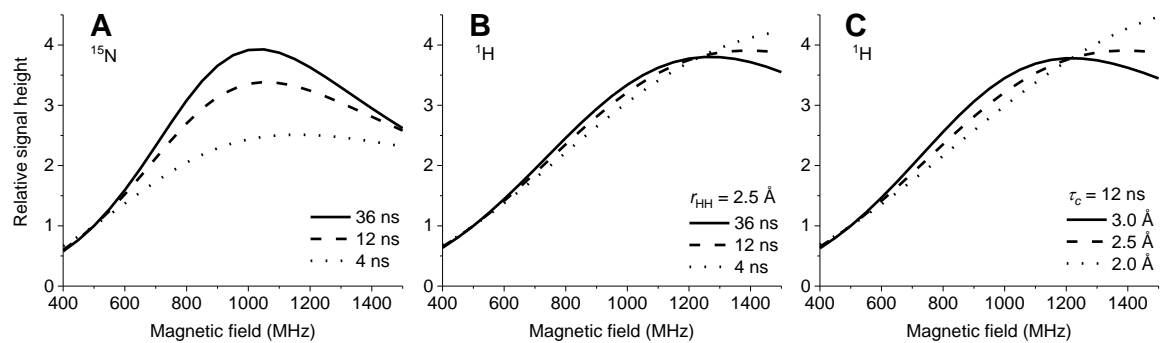

Figure 3. The relative signal intensity of TROSY-HSQC experiments with nitrogen detection (A) and with proton detection (B, C) as a function of the magnetic field taking into account the linewidth, the sensitivity of the acquisition and the signal loss due to transverse relaxation during the INEPT steps. The reported values are scaled by the signal intensity at 500 MHz, and are calculated for an effective proton-proton distance of 2.5 Å and reorientation correlation times of 4, 12, and 36 ns (A, B) or for an effective proton-proton distance of 2.0, 2.5 and 3.0 Å and a reorientation correlation time of 12 ns (C). Other parameters: $\Delta\sigma(^{15}N)$ = -160 ppm, $\Delta\sigma(^{1}H)$ = 16 ppm, $\theta$ = 17° ($\Delta\sigma$ is the difference between the axial and the perpendicular principal components of the axially symmetric chemical shift tensor and $\theta$ is the angle between the tensor axes of the dipole-dipole and CSA interaction).(6, 7)

Therefore, we also compare the nitrogen-detected TROSY spectra acquired at the same fields using the same parameters [number of points and scans (80)] and in the same amount of time(in ~4 hours) (Figures 4,5).

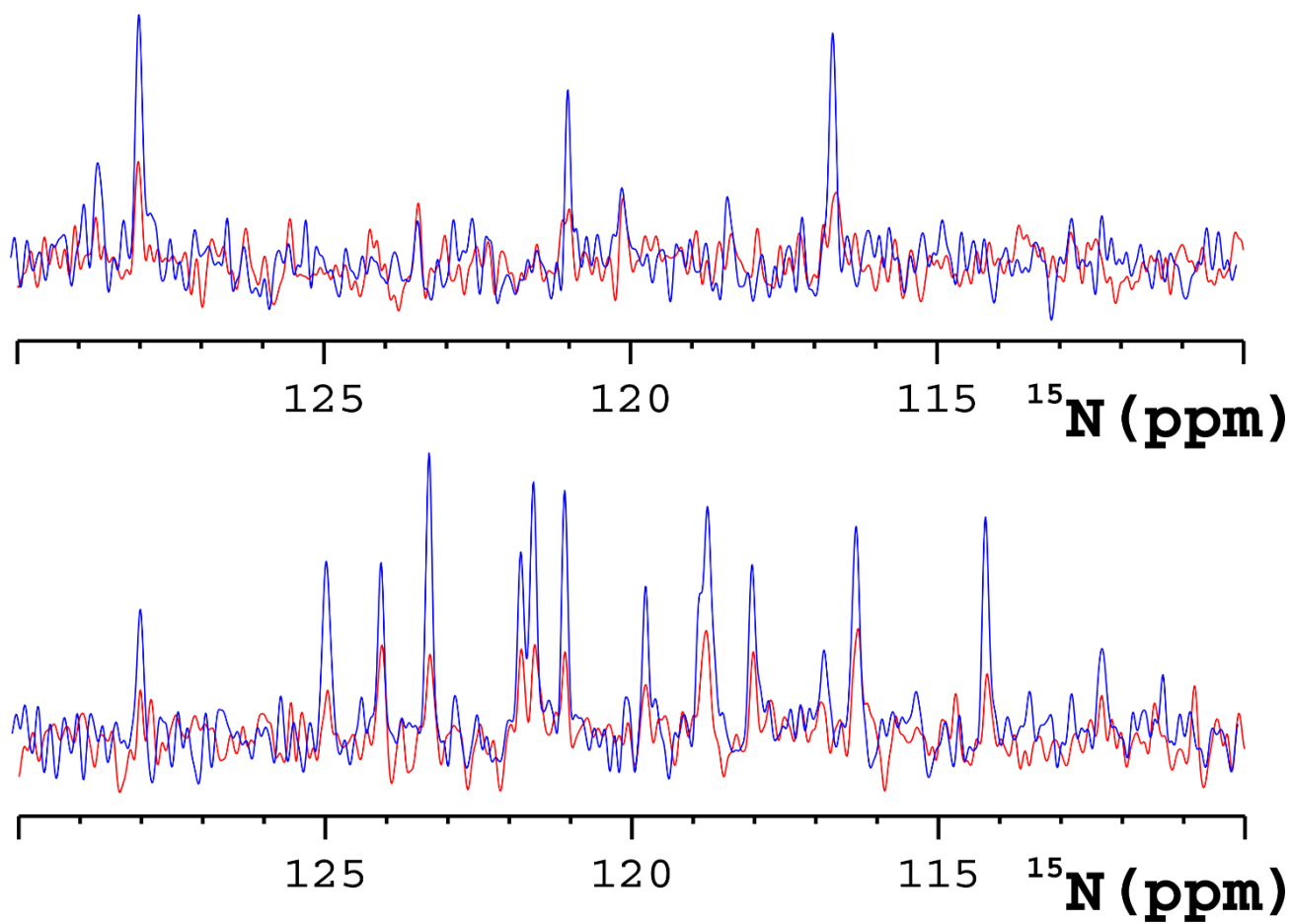

Figure 4. Slices of the $^{15}$N-detected 2D $^{15}$N-$^1$H TROSY spectra [at the $^1$H frequency of 6.49 (top) and 7.38 ppm (bottom)] recorded on 950 Bruker AVANCE III HD (red) and 1200 Bruker AVANCE NEO (blue) spectrometers on a sample of ZnHCAII. The 950 MHz spectrometer is equipped with a 5 mm TCI triple resonance HCN cryo-probehead, while the 1200 MHz spectrometer is equipped with a 3 mm TCI triple resonance HCN cryo-probehead, benefitting from a cryogenically cooled preamplifier for the $^{15}$N channel. The spectra have been processed using the same number of points per ppm and the intensities scaled to equalize the noise level for a better comparison of the signals. The signals appear sharper and more intense in the spectrum acquired at 1200 MHz.

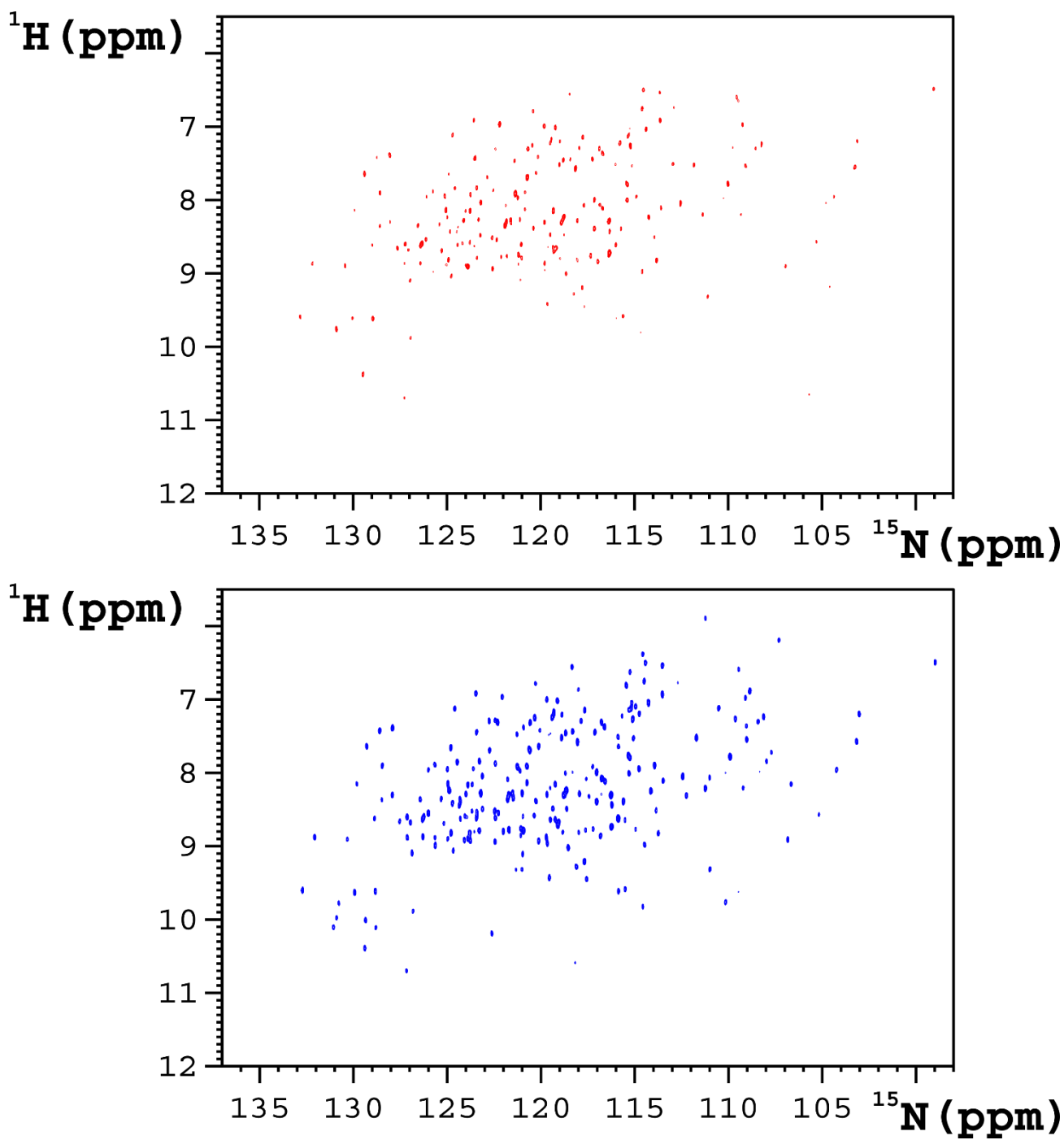

Figure 5. $^{15}$N-detected 2D $^{1}$H-$^{15}$N TROSY-HSQC recorded on 950 Bruker AVANCE III HD (red) and 1200 Bruker AVANCE NEO (blue) spectrometers on a sample of ZnHCAII. The 950 MHz spectrometer is equipped with a 5 mm TCI triple resonance HCN cryo-probehead, while the 1200 MHz spectrometer is equipped with a 3 mm TCI triple resonance HCN cryo-probehead, benefitting from a cryogenically cooled preamplifier for the $^{15}$N channel.

The TROSY effect on a paramagnetic metalloprotein

Given that more relaxation mechanisms are provided by the presence of a paramagnetic metal ion,(8) the interplay of the different mechanisms (9, 10) gives rise to different patterns in the TROSY multiplet (11). This effect is clearly impactful at very high magnetic fields, where the Curie-spin mechanism is highly effective.

In the presence of unpaired electron(s), Curie-spin relaxation can arise due to modulation of the dipole-dipole interaction between the nuclear magnetic moment and the thermal average of the electron magnetic moment, resulting from the small difference in the population of the electron spin levels.(12, 13) The Curie-spin relaxation increases with the square of the magnetic field so that at large fields it can provide contributions even larger than electron-nucleus dipole-dipole relaxation whenever the reorientation correlation time of the molecule is much longer than the electron relaxation time.(14) The cross-correlation between Curie-spin and other (dipole-dipole, CSA) relaxation mechanisms originates differential line broadening in TROSY-type experiments.

In Figure 6, we show some of the signals of cobalt(II)-human carbonic anhydrase II (CoHCAII) where the relative intensity of the peaks in the TROSY multiplet is clearly different from the typical situation in a diamagnetic system (recall that in nitrogen-detected TROSY the sharpest peak in a diamagnetic system is the top left one).

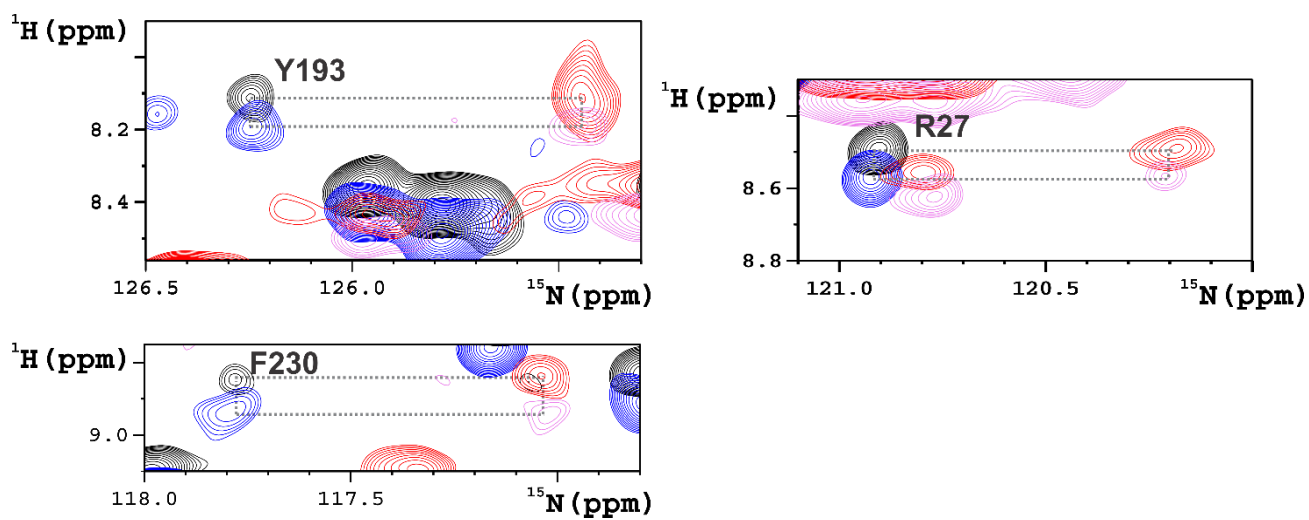

Figure 6. TROSY multiplets of $^{15}$N-detected 2D $^{1}$H-$^{15}$N TROSY-HSQC experiment acquired on CoHCAII at 1200 MHz. The four components of the multiplets are in black, blue, red and pink, respectively; in nitrogen-detected TROSY the sharpest peak in a diamagnetic system would be the top left one (black), while in a paramagnetic system the relative intensities of the components may differ from a residue to another according to NH vector orientations in the tensor frame. Interstingly, the NH bond of Y193 and R27 is almost perpendicular to the NH bond in F230.

Paramagnetic self-alignment

The presence of a paramagnetic center with an anisotropic magnetic susceptibility causes a molecular alignment that, to the first order, is proportional to the square of the magnetic field:(15, 16)

$$S_{ii} = \frac{B_0^2}{15\mu_0 kT}\left(\chi_{ii} - \frac{\chi_{jj} + \chi_{kk}}{2}\right)$$

This is most easily observed in the manifestation of the residual dipolar couplings: given that not all the molecular orientations with respect to the magnetic field are sampled equally, the dipolar couplings (D) among the different nuclei average to a value that is non-zero and depend on the orientation of the internuclear vector in the frame of the magnetic susceptibility anisotropy:

$$\begin{aligned}\mathrm{D} &= -\frac{\mu_0 \hbar \gamma_A \gamma_B}{8\pi^2 r_{AB}^3}\left[S_{zz}(3\cos^2\alpha - 1) + (S_{xx} - S_{yy})\sin^2\alpha \cos 2\beta\right] \\ &= -\frac{B_0^2}{15kT}\frac{\hbar \gamma_A \gamma_B}{8\pi^2 r_{AB}^3}\left[\Delta\chi_{ax}(3\cos^2\alpha - 1) + \frac{3}{2}\Delta\chi_{rh}\sin^2\alpha \cos 2\beta\right]\end{aligned}$$

where $\alpha$ is the angle between the AB internuclear vector and the z-axis of the magnetic susceptibility tensor and $\beta$ is the angle between the x-axis of the tensor and the projection of the internuclear vector in the plane perpendicular to the z-axis.

RDCs can be observed in paramagnetic metalloproteins (17) and are a very important source of structural and dynamical information: (18, 19) there is no dependence of their value on the distance from the metal center, therefore they are intrinsically long range. Furthermore, even a minor change in the backbone orientation can significantly alter the internuclear vector orientation, therefore RDCs are very sensitive to the local geometry. Finally, motions in a wide range of timescales can reduce the expected values of the couplings. Therefore, it is extremely relevant to measure them as precisely and accurately as possible to be able to analyze the structure and the dynamics of a biological macromolecule. The RDCs are usually measured as a variation of the splitting due to J coupling, for example $^1$H and $^{15}$N in an experiment that resolves the splitting (e.g. IPAP (20, 21)): the splitting of the signals is therefore proportional to J+D instead of J only.

As RDCs increase with the square of the field, they are measured with significantly higher precision at higher field (Figure 7). For instance, at 1.2 GHz the RDCs on a particular system are expected to be a factor 1.44 larger than those measured for the same system at 1.0 GHz.

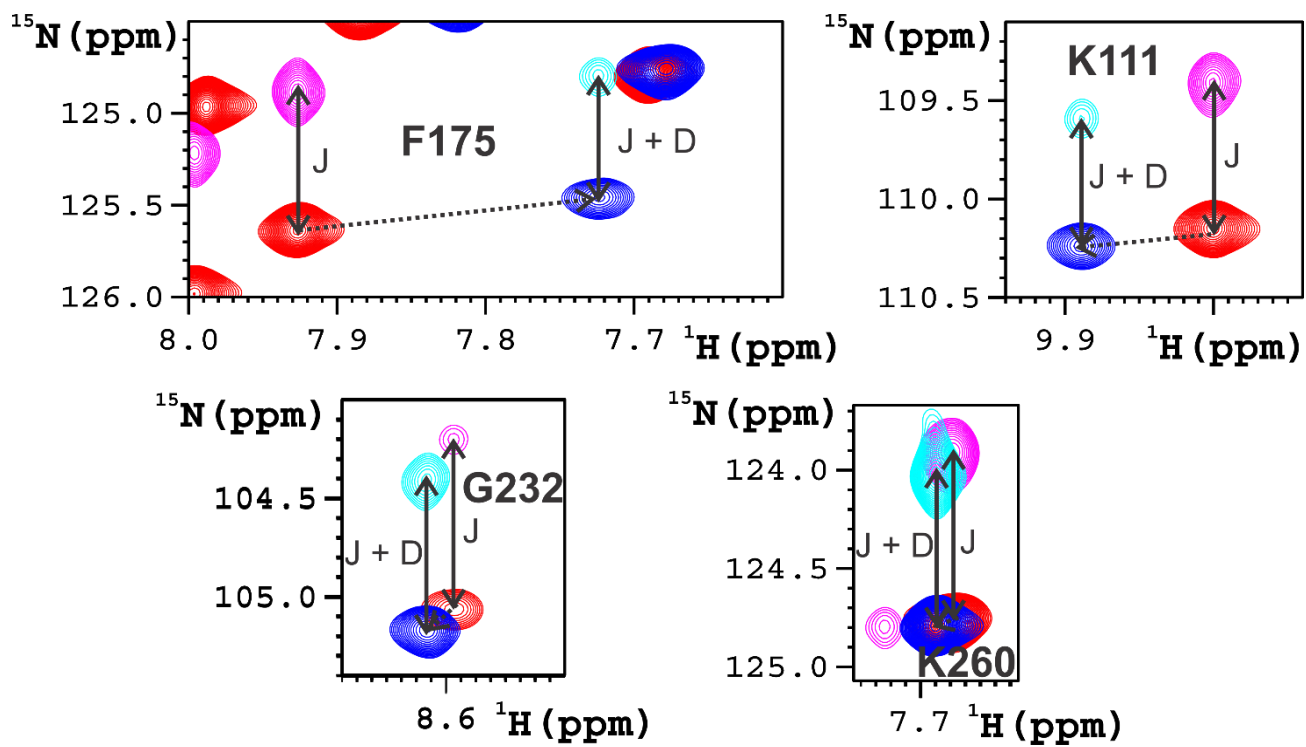

Figure 7. Splitting of four residues (F175, K111, G232 and K260) measured on the ZnHCAII (red and magenta peaks) and CoHCAII (blue and cyan peaks) at 1200 MHz. The measured RDCs are positive for F175 and K111, while negative for G232 and K260.

Heteronuclear detection and linewidths in intrinsically disordered proteins

Intrinsically disordered proteins (IDPs) are involved in a wide range of biological processes, both physiological and pathological. From a structural standpoint, IDPs are polypeptide chains exhibiting limited or no long-range ordering, and they usually contain repetitive sequence elements. These two features contribute to the lack of the chemical shift dispersion that is usually associated with folded proteins.

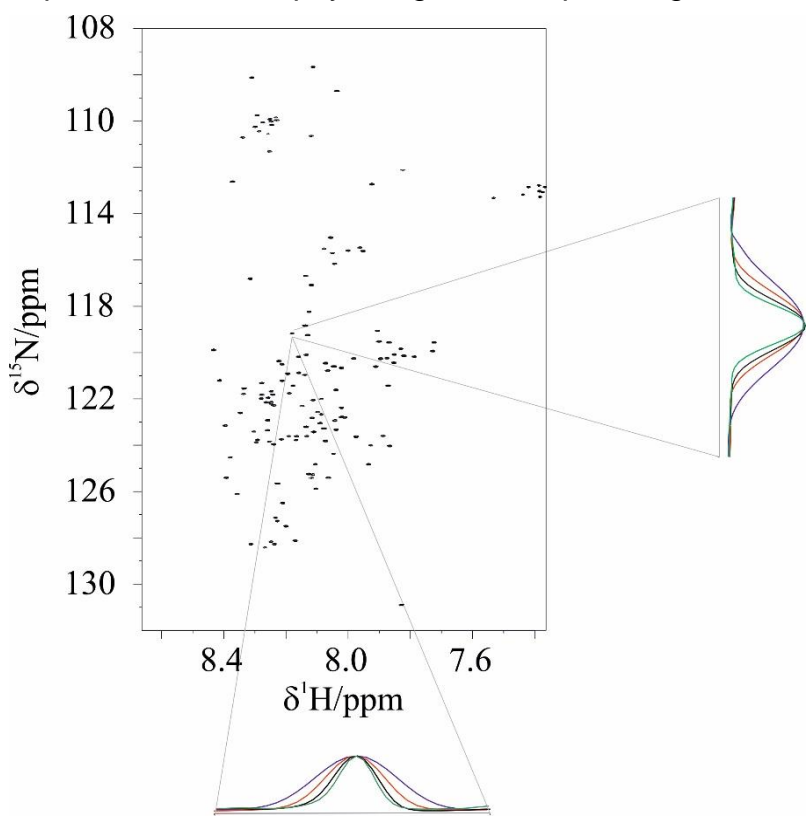

Figure 8. $^1$H-$^{15}$N HSQC spectra of □-synuclein at 1200 MHz. For one peak (S129) we report the linewidths at different fields, which decrease sizeably when expressed in ppm, from 500 MHz (blue), 700 MHz (red), 950 MHz (black) and 1200 MHz (green).

At the same time, their unique mobility pattern makes their peaks appear sharp also at very high fields.(22-24) Therefore, the resolution of the spectra is expected to grow significantly upon moving to higher fields (Figure 8).

One strategy for improving the spectral resolution is to move from $^1$H detection to the detection of heteronuclei, which intrinsically have wider chemical shift dispersions despite the lack of structure. The first choice is $^{13}$C detection, which results feasible even at fields as low as 700 MHz.(25, 26) The spectra acquired at 1.2 GHz have a remarkably high resolution (Figure 9).

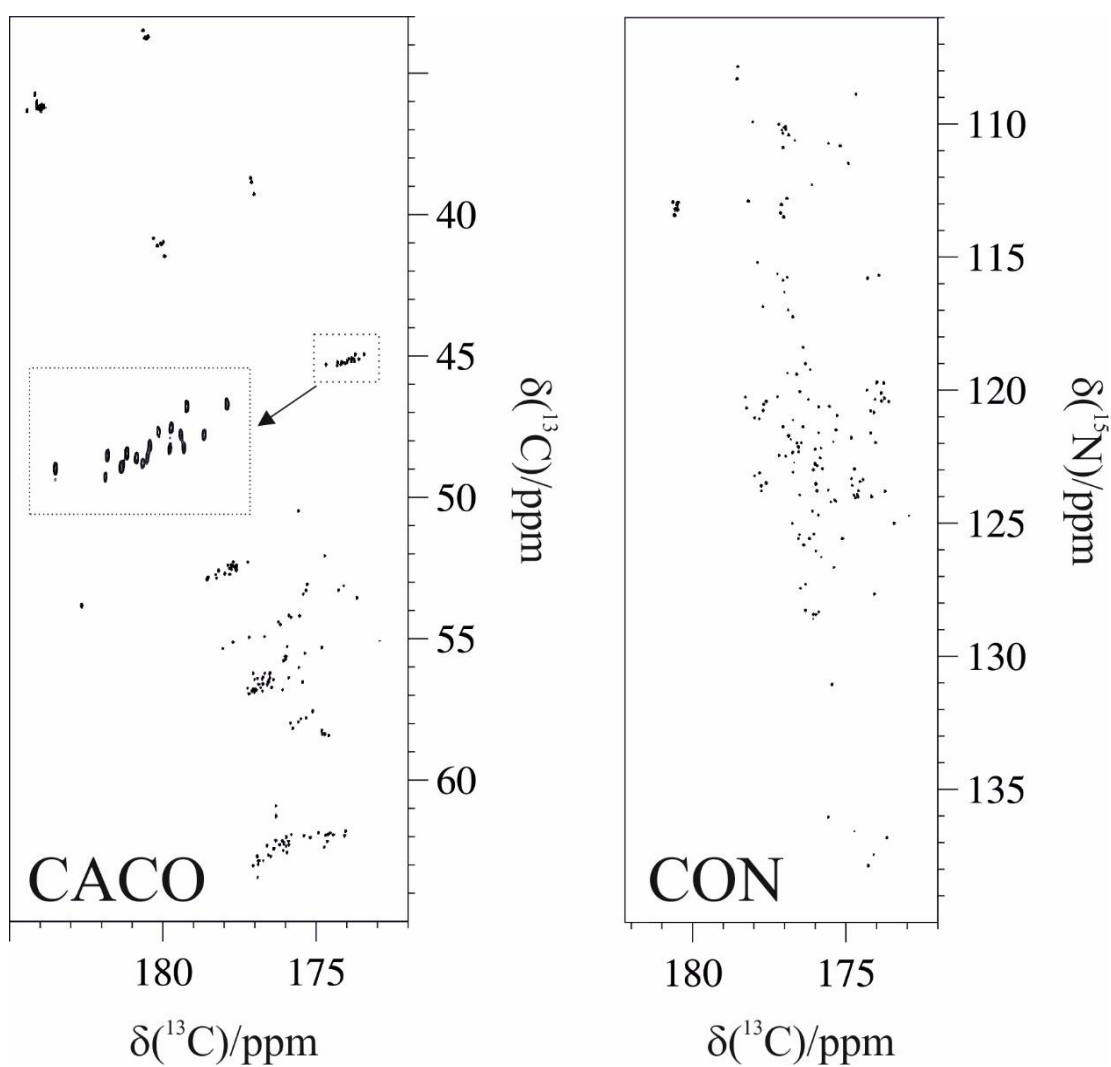

Figure 9. CACO and CON spectra recorded on α-synuclein at 1.2 GHz. In the inset, the region where the glycine residues signals resonate is shown to demonstrate the remarkable resolution that can be achieved.

As for folded proteins, the possibility of detecting nitrogen resonances offers the possibility of achieving an unprecedented resolution (Figure 10).

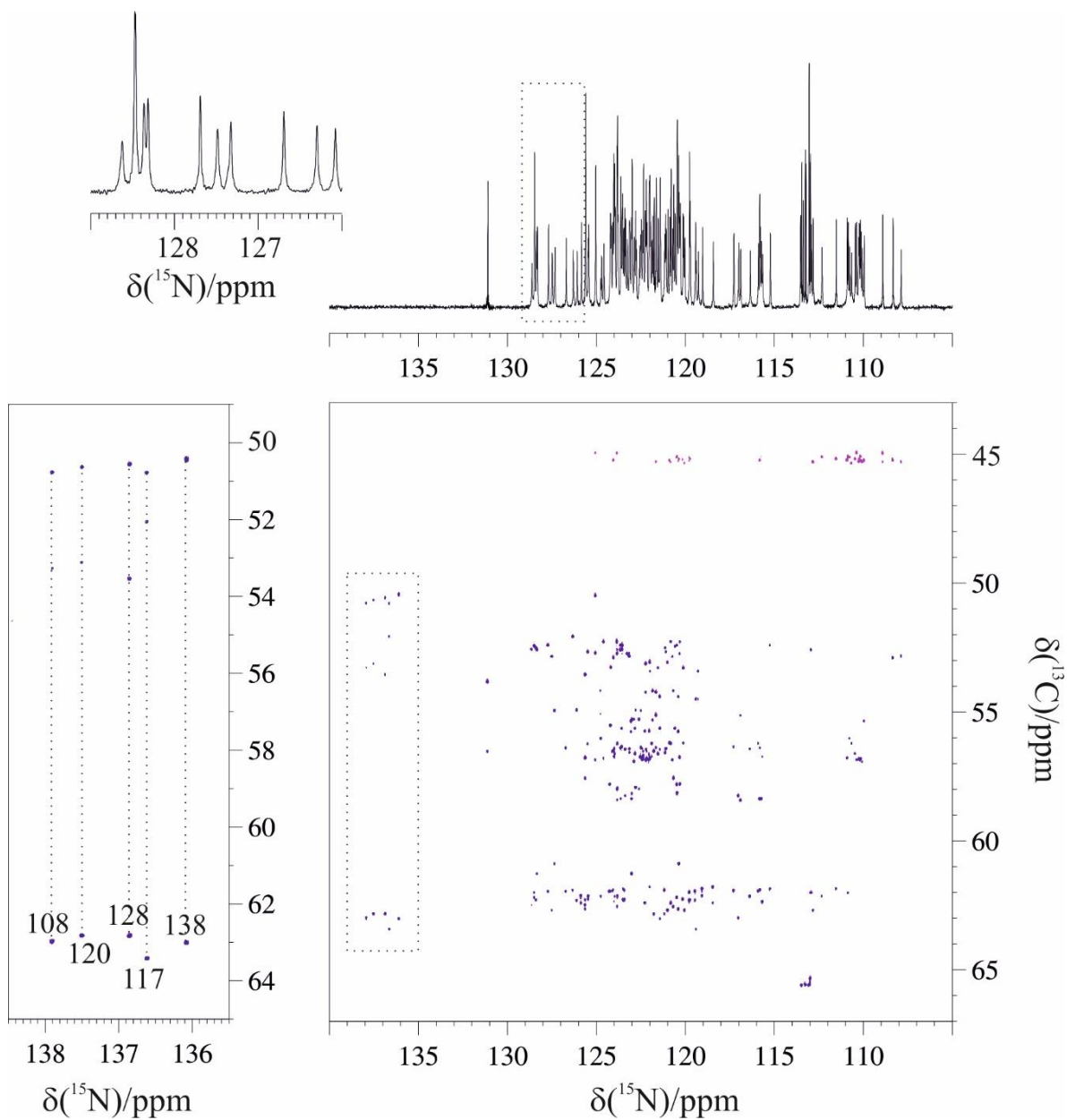

Figure 10. $^{15}$N detected spectra recorded on α-synuclein at 1.2 GHz. The $^{15}$N 1D spectrum is shown on top ($^{1}$H-$^{15}$N refocused INEPT variant); the panel on the left reports an enlargement of one region (dotted box) to show the excellent resolution of $^{15}$N resonances. The 2D (H)CAN is shown on the bottom; an enlargement of the proline region (dotted box) is reported on the left with the sequence specific assignment to the five proline residues of α-synuclein.

Sensitivity and resolution in natural abundance systems

The assessment of the "Higher Order Structure" (HOS) by NMR is emerging as the technique of choice to control the manufacturing process and for comparability studies of biologics.(27) When size and concentration of the protein prevent the use of $^1$H-$^{15}$N correlation spectroscopy, $^{13}$C methyl NMR can be successfully applied. As already mentioned, the overall sensitivity of NMR increases with increasing magnetic field. This allows for the detection of proteins at low concentrations and with e.g. $^{13}$C in natural abundance.

The most striking feature of the spectra acquired at the higher field is the higher resolution of the methyl resonances. As an example, we report two alternate SOFAST-HMQC (ALSOFAST-HMQC) spectra (Figure 11) collected on a non-glycosylated antibody of 145 kDa at 900 MHz and 1.2 GHz, respectively, with comparable acquisition times (total experimental time is 4 h 30' at 1.2 GHz and 5 h 40' at 900 MHz). The resolution enhancement achieved at 1.2 GHz can be beneficial in the assessment of HOS and in comparability studies.

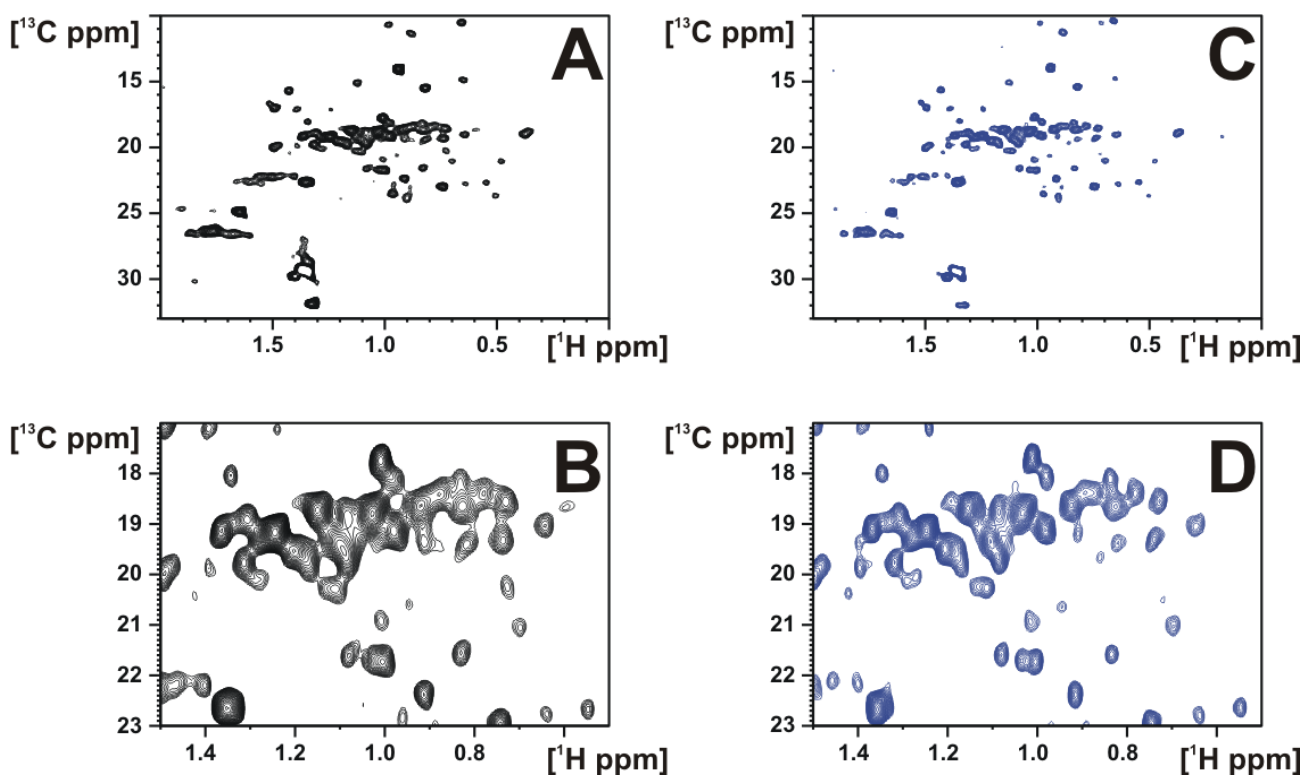

Figure 11. 2D ALSOFAST-HMQC spectra collected on a non-glycosylated antibody of 145 kDa at 313 K. The spectra collected on a magnet operating at 900 MHz are reported in panels A and B. The spectra collected on the magnet operating at 1.2 GHz are reported in panels C and D.

Sensitivity and resolution in diluted systems: a perspective for in-vivo applications

The increased sensitivity of the 1.2 GHz is extremely beneficial to characterize proteins withtin human cells by in-cell NMR.(28, 29) Indeed, cell samples are often limited in quantity, and/or suffer from high physiological ionic strength, and must be analyzed in small volumes (e.g. 3 mm or 4 mm shigemi tubes), thus sacrificing sensitivity. Such improvement will allow detection of intracellular proteins at lower concentrations, closer to the physiological levels. An additional benefit from the higher field comes from the slowed down chemical/conformational exchange regime of certain nuclei, such as amide protons in proteins. Living cells must be analyzed at physiological pH and temperature, where such exchange phenomena are exacerbated. Spectra collected at 1.2 GHz on a diluted globular protein (human carbonic anhydrase II, 100 µM) show that imino protons from zinc-binding histidines, that undergo solvent exchange in the slow-intermediate regime, are detected with excellent S/N ratio (Figure 12).

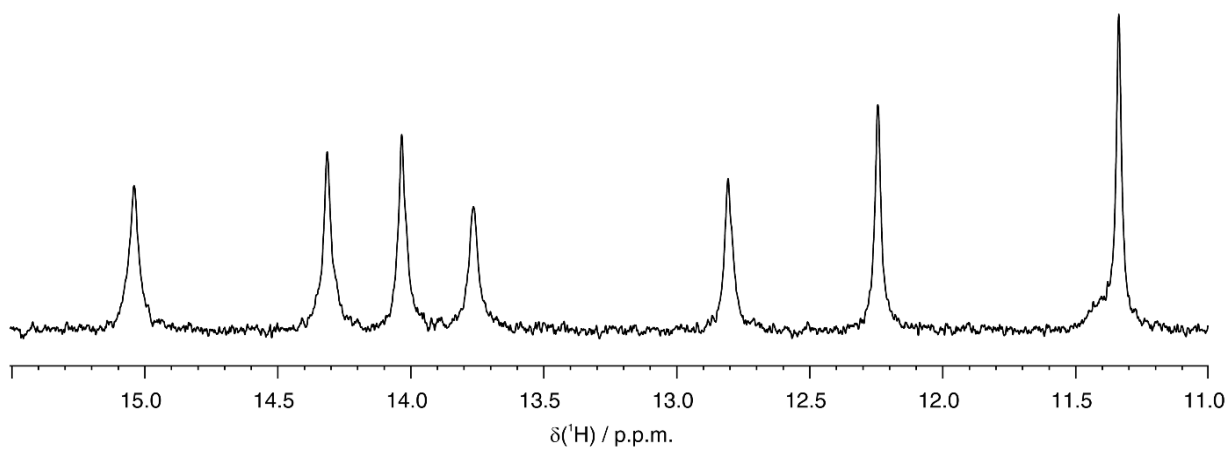

Figure 12. Imino region of the $^1$H NMR spectrum acquired at 310 K on a sample of carbonic anhydrase II (100 µM). A binomial water suppression WATERGATE scheme was employed (p3919gp).

Sensitivity and resolution in metabolomics

Metabolomics, i.e. 'omic' science devoted to the study of the complex set of metabolites present in living organisms,(30) is, like other 'omic' sciences, a discipline driven by technological advancements. Its evolution takes advantage of new improvements in instrumentation, analytical techniques, statistical methods and software to accelerate or improve the collection, the analysis, and the interpretation of the data. (31) NMR is a main analytical tool in metabolomics (32). It is highly reproducible and automatable, thus permitting high-throughput, large-scale metabolomics studies.(33) However, NMR also has few disadvantages, the most significant being its lack of sensitivity and, especially in crowded spectra of complex biofluids (more than 2000 metabolites are estimated in urine), lack of resolution. These drawbacks are closely linked to the magnetic field strength employed, thus in principle can be mitigated using high- or ultrahigh-field NMR spectrometers. These machines offer better sensitivity with higher resolution, allowing the identification and quantification of more metabolites with respect to the common magnetic fields used in metabolomics (400-700 MHz with 600 MHz being the standard).While high throughput routine metabolomics studies are better performed at intermediate fields (with 600 MHz being the standard instruments), the availability of higher and higher fields can help in extending the assignment of more and more metabolites, in constructing more and more informative databases, and in developing more and more accurate assignment tools.(34) The first 1.2 GHz NMR spectrum of a urine sample is reported in Figure 13.

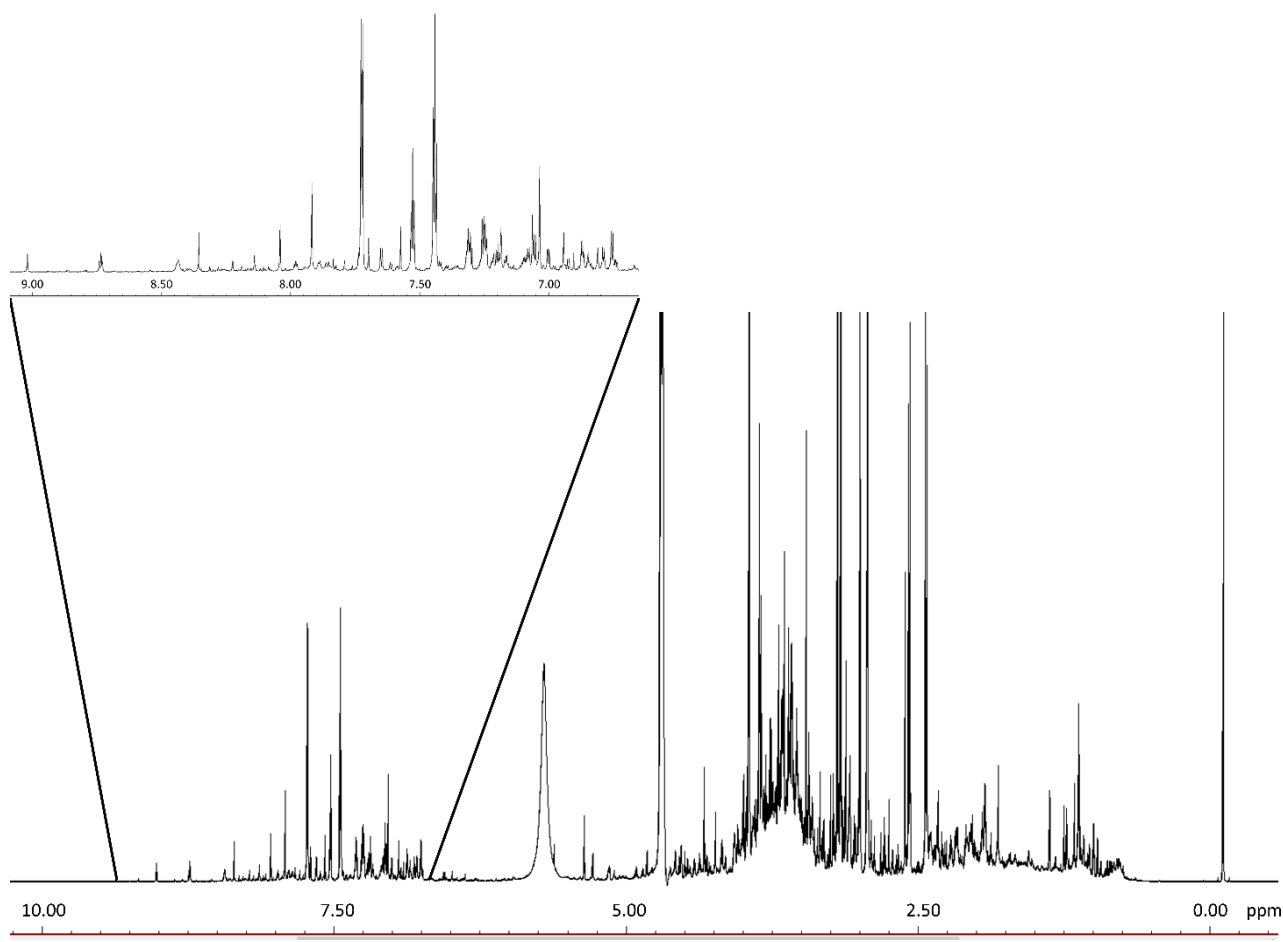

Figure 13. 1.2 GHz NMR spectrum of a human urine sample. The aromatic region is expanded for clarity.

Acknowledgements. We acknowledge Rainer Kuemmerle and Helena Kovacs for the precious help, the competent assistance and the stimulating discussions during the visit of